\documentclass[aps,showpacs,twocolumn]{revtex4}
\usepackage{epsfig}
\usepackage{slashed}
\usepackage{color}

\begin{document}

\title{The contributions of the vector-channel at finite isospin chemical potential with the self-consistent mean field approximation}

\author{Zu-Qing Wu$^a$}\email{wujieyi1001@foxmail.com}
\author{Chao-Shi$^b$}\email{shichao0820@gmail.com}
\author{Jia-Lun Ping$^c$}\email{jlping@njnu.edu.cn}
\author{Hong-Shi Zong$^{d,e,f}$}\email{zonghs@nju.edu.cn}

\affiliation{$^a$Department of Physics, Nanjing University, Nanjing 210093, P. R. China}

\affiliation{$^b$Department of nuclear science and technology, Nanjing University of Aeronautics and Astronautics, Nanjing 210016, P. R. China}

\affiliation{$^c$Department of Physics, Nanjing Normal University, Nanjing 210023, P. R. China}

\affiliation{$^d$Department of Physics, Nanjing University, Nanjing 210093, P. R. China}

\affiliation{$^e$Nanjing Proton Research and Design Center, Nanjing 210093, P. R. China}

\affiliation{$^f$Department of Physics, Anhui Normal University, Wuhu 241000, P. R. China}

\begin{abstract}
The self-consistent mean field approximation of two-flavor NJL model, which introduces a free parameter $\alpha$
($\alpha$ reflects the weight of different interaction channels), is employed to investigate the contributions of
the vector-channel at finite isospin chemical potential $\mu_I$ and zero baryon chemical potential $\mu_B$ and zero temperature $T$.
The calculations show that the consideration of the vector-channel contributions leads to lower value of pion condensate
in superfluid phase, compared with the standard Lagrangian of NJL model ($\alpha=0$). In superfluid phase,
we also obtain lower isospin number density, and the discrepancy is getting larger with the increase of isospin potential.
Compared with the recent results from Lattice QCD, the isospin density and energy density we obtained with $\alpha=0.5$
agree with the data of lattice well. In the phase diagram in the $T-\mu_I$ plane for
$\mu_B=0$, we can see that the difference of the critical temperatures of phase transition between the results with $\alpha=0$
and $\alpha=0.5$ is up to $3\%-5\%$ for a fixed isospin potential. All of these indicate that the vector channels play an
important role in isospin medium.
\end{abstract}


\maketitle

\section {Introduction}
The study of thermodynamics of strongly interacting system under extreme conditions is helpful for us to
develop better understanding to the physical scene shortly after the Big Bang \cite{Boyanovsky,Pizzone},
the structure of compact stars \cite{Bielich,Weber} and heavy-ion collision experiments \cite{Marty,Xiaofeng}.
The probe of the properties of strongly interacting matter at such large temperatures and densities is
carried on at CERN, BNL and GSI, etc.\cite{Tannenbaum,Schmidt}, in particular searching for the position
or even the existence of the critical end point (CEP) \cite{XiaofengLuo}.

One of the interesting phenomena in extending the phase diagram to nonzero isospin region is the appearance
of a new phase. It is known that the isospin chemical potential has an effect on hadronic matter, which can
rotate the quark-antiquark condensate, and the phenomenon is named as pion condensation, since it indicates
the direction of the $U_I(1)$ symmetry breaking corresponding to the conservation of the pion number. When
the isospin chemical potential exceeds the pion mass ($\mu_I>m_\pi$), a superfluid of charged pions in the zero momentum state
will occur, i.e. the pion superfluid phase. This Bose-Einstein condensate (BEC) of pions is an electromagnetic
superconductor \cite{Stefano,D.T.Son}. Different from the normal phase ($\mu_I \le m_\pi$) where pion condensation
is zero, the realization of pion condensation can change the low energy properties of matter, such as the
mass spectrum and the lifetimes of mesons \cite{Andrea,Migdal,Kogut}, and are also related to a lot of
phenomena \cite{B,Viktor}. Therefore, it is important for us to study the thermodynamics of strongly interacting
system in an isospin medium .

Theoretically, Quantum Chromodynamics (QCD) is recognized as the fundamental theory of the strong interaction.
It is generally thought that there exist rich phase structures of Quantum Chromodynamics at finite temperature and finite
density. At high temperatures and/or high densities, the perturbative QCD can describe the nature of the phases well.
At vanishing density and finite temperature, lattice simulations from first principle of QCD have provided valuable insights
into the QCD phase diagram. Nevertheless, at finite baryon density lattice simulations are hindered by the sign
problem \cite{Karsch} and there is no problem for lattice simulations at finite isospin density in principle \cite{Son}.
Beside that, there are lots of low-energy effective models, such as chiral perturbation theory \cite{Loewe,Kogut}, random matrix
method \cite{Klein,Arai} and Nambu--Jona-Lasinio (NJL) model \cite{Barducci,He,Tao,Lian}, to be used as tools to investigate
the phase structures in isospin matter, in which NJL model described the chiral dynamics of QCD well \cite{S.P.Klevansky}.

In this paper, we study the contributions of the vector channels at finite isospin potential in the framework of NJL model
with the self-consistent mean field approximation. The standard Lagrangian of NJL model contains scalar $\left(\bar{\psi}\psi\right)^2$
and pseudoscalar-isovector $\left(\bar{\psi}i\gamma_5\mbox{\boldmath{$\tau$}}\psi\right)^2$ channels \cite{S.P.Klevansky}.
Making use of Fierz transformations, not only scalar and pseudoscalar-isovector channels but also other interaction channels
can be produced. These interaction channels play an important role in the case of external fields. For instance, when the finite chemical
potential is involved, the vector-isoscalar channel is very important \cite{Walecka}. Similarly when we discuss the axial chemical
potential, the isovector-isoscalar channel becomes very important \cite{S,BW,YL,ZF}, and in the study of the chirally imbalanced system,
the contribution of the axial-vector channel cannot be neglected \cite{Yang}. In this connection, if we study the system under the condition of finite isospin density, the contributions of the vector-isoscalar channels and the pseudoscalar-isovector channels should be considered.
In previous analyses of NJL model, people usually ignore the contributions of various channels from the Fierz-transformed term or
manually add the relevant terms \cite{Kunihiro}. As shown below, the above mean field approximation approach is not self-consistent.
In this paper, we will employ the self-consistent mean field approximation \cite{FeiWang} of NJL model to study the thermodynamics of
strongly interacting system at nonzero isospin chemical potential. This model introduces a free parameter $\alpha$ to reflect the
proportion of the different channel contributions from the Fierz-transformed term.

This paper is organized as follows. In Sec. \ref{s2} the two-flavor NJL model by the self-consistent mean field approximation
in the case of isospin chemical potential is introduced and we can get the self-consistent gap equations. In Sec. \ref{s3}, we discuss
the contributions of the vector channels to relevant thermodynamic quantities and phase diagram, and the comparison of our results with
lattice data is also given. In the last section, we summarize our findings.

\section {The self-consistent mean field approximation of NJL model}
\label{s2}
In the present work, only two flavors are considered, i.e. $N_f=2$. The standard flavor $SU(2)$ NJL model
Lagrangian density is defined as \cite{S.P.Klevansky}
\begin{equation}
{\cal L}_{NJL} =
\bar{\psi}\left(i\slashed{\partial}-m_0\right)\psi
+G\left[\left(\bar{\psi}\psi\right)^2+\left(\bar{\psi}i\gamma_5\mbox{\boldmath{$\tau$}}\psi\right)^2
\right]
\end{equation}
with scalar and pseudoscalar interactions related to $\sigma$ and ${\bf \pi}$ excitations respectively, where $G$ is the coupling constant and the matrix of the current quark mass is $m_0 = \text{diag}(m_{0u},m_{0d})$.

Performing the Fierz transformation \cite{S.P.Klevansky} on the four-Fermion interaction terms, we have
\begin{eqnarray}
&{\cal L}_{IF}& =
\frac{G}{8N_c} \left[2\left(\bar{\psi}\psi\right)^2+2\left(\bar{\psi}i\gamma_5\mbox{\boldmath{$\tau$}}\psi\right)^2
-2\left(\bar{\psi}\mbox{\boldmath{$\tau$}}\psi\right)^2 \right. \nonumber\\
&& -2\left(\bar{\psi}i\gamma_5\psi\right)^2-4\left(\bar{\psi}\gamma^{\mu}\psi\right)^2-4\left(\bar{\psi}i\gamma^{\mu}\gamma_5\psi\right)^2
 \nonumber\\
&&\left.+\left(\bar{\psi}\sigma^{\mu\nu}\psi\right)^2 -\left(\bar{\psi}\sigma^{\mu\nu}\mbox{\boldmath{$\tau$}}\psi\right)^2
\right],
\end{eqnarray}
where color octet contributions have been neglected and the number of colors is $N_c=3$. Then the Lagrangian becomes
\begin{equation}
{\cal L}_F =
\bar{\psi}\left(i\slashed{\partial}-m_0\right)\psi+{\cal L}_{IF}.
\end{equation}
The original Lagrangian ${\cal L}_{NJL}$ and the transformed Lagrangian ${\cal L}_{F}$ are equivalent, since the Fierz transformation
is a mathematical identity transformation. Considering the ${\cal L}_{NJL}$ and ${\cal L}_{F}$ are mathematically equivalent, the most
general effective Lagrangian can be introduced \cite{FeiWang}: ${\cal L}_R = (1-\alpha) {\cal L}_{NJL} + \alpha {\cal L}_F$ where
the parameter $\alpha$ is an arbitrary $c$-number, and the Lagrangian does not change with $\alpha$. Through Fierz transformation,
we can obtain more general interaction terms. It is very helpful for us to understand and deal with the problem of strongly interacting
system under the condition of external fields. As mentioned in the introduction above, the various channel contributions are not negligible
if we study the system with external fields.

However, when the mean field approximation is applied, the contributions of ${\cal L}_{NJL}$ and ${\cal L}_{F}$ are found to be no longer
identical, this is because the Fierz transformation and the mean field approximation are not commutative. Especially when
the system is in external fields, the results yielded by two Lagrangians are very different \cite{S.P.Klevansky}. This means that it is
important for us to know the contributions of each interacting term once the mean field approximation is used. Actually, just as pointed out by Refs. \cite{FeiWang,TongZhao,Q}, there is no physical requirement to constrain the value of $\alpha$. $\alpha$ in principle should
be determined by experiments rather than the self-consistent mean field approximation itself. The Lagrangian of the self-consistent mean field
approximation is adopted as $\langle{\cal L}_R\rangle_m = \left(1-\alpha\right) \langle{\cal L}_{NJL}\rangle_m + \alpha \langle{\cal L}_F\rangle_m$
\cite{FeiWang}, where $\langle ¡­\rangle_m$ denotes the mean field approximation.

In order to investigate the system where $u$- and $d$- quark are asymmetric, we can introduce the isospin chemical potential $\mu_I$ which
connects to the isospin number density $n_I=(n_u-n_d)/2$. Just as the chemical potential $\mu$ can reflect the density $n$ of the quark,
we introduce the isospin chemical potential to denote the imbalance between the $u$- and $d$-quarks. In the imaginary time formulism of finite
temperature field theory \cite{J.I.Kapusta}, the partition function for a system at finite baryon and isospin densities can be represented as
\begin{equation}
\label{s2 40}
Z\left(T,\mu_I,\mu_B,V\right) = \int \left[d\bar{\psi}\right]\left[d\psi\right]
  e^{\int_{0}^{\beta} d\mbox{\boldmath{$\tau$}} \int d^3\vec{x} \left({\cal L}
 + \bar{\psi}\mu\gamma_0\psi\right)}\ ,
\end{equation}
where $V$ is the volume of the system, $\beta$ is the inverse temperature $\beta = 1/T$; the quark number density operator $\hat{n}$ and the quark isospin number density operator $\hat{n}_I$ are $\bar{\psi}\gamma_0\psi$ and ${1 \over 2}\bar{\psi}\gamma_0\tau_3\psi$, and $\mu_B$ and $\mu_I$ are the baryon and isospin
chemical potentials, where $\mu = \text{diag}(\mu_u,\mu_d)$ is the matrix of quark chemical potential in flavor space with the $u$
and $d$ quark chemical potentials,
\begin{eqnarray}
\label{s2 2}
\mu_u &=& \frac{\mu_B}{3}+\frac{\mu_I}{2}\ ,\nonumber\\
\mu_d &=& \frac{\mu_B}{3}-\frac{\mu_I}{2}\ ,
\end{eqnarray}
the factors $\frac{1}{3}$ and $\frac{1}{2}$ reflect the fact that $3$ quarks make up a baryon and quark's isospin quantum number is $\frac{1}{2}$.

Following Refs. \cite{FeiWang,TongZhao,Q}, the equivalent Lagrangian in this work can be rewritten as
\begin{equation}
{\cal L}_R= (1-\alpha) {\cal L}_{NJL} + \alpha {\cal L}_F +  \bar{\psi}\textcolor[rgb]{0.00,0.50,1.00}{\mu}\gamma_0\psi.
\end{equation}

In our study, we only care about the contributions from scalar, vector and pseudoscalar-isovector channels. Other terms have no effect on
our calculation at the level of mean field approximation. Applying the mean field approximation to this Lagrangian and dropping the irrelevant
terms, we can get the effective Lagrangian
\begin{eqnarray}
\label{s2 1}
{\cal L}_{eff}
&=&\bar{\psi}\left(i\slashed{\partial}-M+\mu^{\prime} \gamma_0
+2G\pi i\gamma_5\tau_1\right)\psi \nonumber\\
&-& G\left(\sigma^2 + \pi^2\right) +\beta n^2\ ,
\end{eqnarray}
where M is called constituent quark mass:
\begin{equation}
\label{s2 8}
M=m_0-2G\sigma,
\end{equation}
and
\begin{equation}
\label{s2 3}
\mu^{\prime} = \mu - 2 \beta n.
\end{equation}

Employing the Eqs. (\ref{s2 2}) and (\ref{s2 3}), one can obtain the following relation at $\mu_B = 0$ (in this paper, only the zero baryon
density case is considered),
\begin{equation}
\label{s2 9}
\mu^{\prime}_I = \mu_I - 8 \beta n_I,
\end{equation}
where for convenience we redefine the parameter $\beta=\frac{-2 G \alpha}{11 \alpha-12}$ in the formalism.

The quark condensation $\sigma=\langle\bar{\psi}\psi\rangle$, the pion condensation
$\pi=\langle\bar{u} i\gamma_5 d\rangle+\langle\bar{d} i\gamma_5 u\rangle$, the quark number density
$n=\langle\bar{\psi}\gamma_0\psi\rangle=\langle\bar{u}\gamma_0u\rangle+\langle\bar{d}\gamma_0d\rangle$ and the isospin number density
$n_I={1 \over 2}\langle\bar{\psi}\gamma_0\tau_3\psi\rangle=(\langle\bar{u}\gamma_0u\rangle-\langle\bar{d}\gamma_0d\rangle)/2$ can be determined in a thermodynamically self-consistent way.
We can insert the effective Lagrangian (\ref{s2 1}) into the partition function (\ref{s2 40}) to get the mean-field thermodynamic potential
\begin{eqnarray}
\Omega &=& -\frac{T}{V} \ln{Z} \nonumber\\
&=& G\left(\sigma^2 + \pi^2\right) - \beta n^2 + \Omega_M ,
\end{eqnarray}
where $\Omega_M$ is expressed as
\begin{eqnarray}
&& \Omega_M = -2N_c \int_{0}^{\Lambda} {d^3\vec{p}\over (2\pi)^3} \left[E_p^- + E_p^+ + 2N_fT\left(\ln\left(1 \right.\right.\right.\nonumber\\
&& \left.\left.\left. +\exp{(-E_p^-/T)}\right) + \ln\left(1+\exp{(-E_p^+/T)}\right)\right)\right],
\end{eqnarray}
here the effective quark energies $E^\pm_p$ are given by
\begin{eqnarray}
E^\pm_p &=& \sqrt{\left(E_p \pm \mu^{\prime}_I/2\right)^2 + 4G^2\pi^2} ,\\
E_p &=& \sqrt{|\mathbf{p}|^2 + M^2}\ .
\end{eqnarray}

Given the extremum condition of the thermodynamic potential $\frac{\partial{\Omega}}{\partial{\sigma}}=0,\frac{\partial{\Omega}}{\partial{\pi}}=0,\frac{\partial{\Omega}}{\partial{n}}=0
,\frac{\partial{\Omega}}{\partial{n_I}}=0$, we can get the quark condensate,
\begin{eqnarray}
\label{s2 4}
\sigma &=& \int_{0}^{\Lambda} {d^3 p \over (2\pi)^3} \frac{2N_cM}{E_p} \left[\frac{E_p-\mu^{\prime}_I/2}{E_p^-}(f(E_p^-)-f(-E_p^-)) \right.\ \nonumber\\
&+& \left. \frac{E_p+\mu^{\prime}_I/2}{E_p^+}(f(E_p^+)-f(-E_p^+))\right]\ ,
\end{eqnarray}
the pion condensate,
\begin{eqnarray}
\label{s2 5}
\pi    &=& -4N_cG\pi \int_{0}^{\Lambda} {d^3 \vec{p}\over (2\pi)^3} \left[\frac{1}{E_p^-} \left(f(E_p^-)-f(-E_p^-)\right) \right.\ \nonumber\\
&+& \left. \frac{1}{E_p^+} \left(f(E_p^+)-f(-E_p^+)\right)\right]\ ,
\end{eqnarray}
the quark number density,
\begin{eqnarray}
\label{s2 6}
n      &=& {2 \over 3} N_c\int_{0}^{\Lambda} {d^3 {p} \over (2\pi)^3} \left[f(E_p^-)+f(-E_p^-)+f(E_p^+) \right.\ \nonumber\\
&+& \left. f(-E_p^+)-2\right]\ ,
\end{eqnarray}
and isospin number density,
\begin{eqnarray}
\label{s2 7}
n_I    &=& N_c\int_{0}^{\Lambda} {d^3 {p} \over (2\pi)^3} \left[ \frac{E_p-\mu^{\prime}_I/2}{E_p^-}(f(E_p^-)-f(-E_p^-)) \right.\ \nonumber\\
&-& \left. \frac{E_p+\mu^{\prime}_I/2}{E_p^+}(f(E_p^+)-f(-E_p^+))\right]\
\end{eqnarray}
with the Fermi-Dirac distribution function
\begin{equation}
f(x) = {1\over e^{x/T}+1}.
\end{equation}
Finally, inserting Eqs. (\ref{s2 4}-\ref{s2 7}) into the Eqs. (\ref{s2 8}-\ref{s2 9}), we will obtain the self-consistent gap equations
in the case of finite isospin chemical potential $\mu_I$.

The parameter set used for the purpose of the present study are the current quark mass $m_{0u}=m_{0d}=m_0=4.76$ $\text{MeV}$,
the coupling constant $G = 4.78\times 10^{-6} \text{MeV}^{-2}$ and the cut-off $\Lambda = 659$ $\text{MeV}$, which are obtained by fitting
the pion mass $m_{\pi}=131.7$ $\text{MeV}$ as used by Lattice QCD \cite{Brandt} at $T = \mu_I = \mu_B = 0$, and other parameters are
the decay constant $f_{\pi}=92.4$ $\text{MeV}$ and the quark condensate per flavor $\langle\bar{\psi}\psi\rangle=-(250~\text{MeV})^3$.

\section {Numerical results and discussion}
\label{s3}
It is mentioned above that the Refs. \cite{FeiWang,TongZhao,Q} indicate the parameter $\alpha$ can be constrained by experiments. One choice,
for example in Ref. \cite{TongZhao,Q}, is that $\alpha$ can be determined by astronomical observation data on the latest neutron star mergering.
However, with the lack of reliable experiment data on the strongly interacting matter at finite density currently, so in our study we consider
$\alpha$ as a free parameter. In this paper, we will show our results with different $\alpha$'s, $\alpha$=0 represents the standard NJL
model \cite{S.P.Klevansky}, $\alpha$=0.5 is found to be in good agreement with recent lattice data, $\alpha$=0.9 is adopted from
Ref. \cite{TongZhao}, and $\alpha$=1.044 is taken from Ref. \cite{FeiWang}.

\begin{figure}
\includegraphics[width=3.5in]{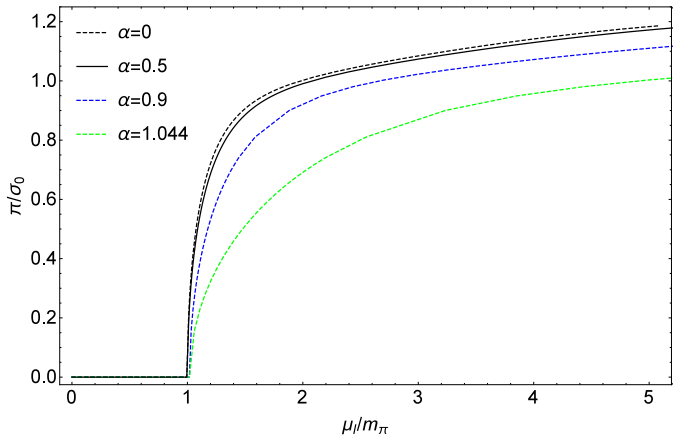}
\caption{The normalized pion condensate $\pi/\sigma_0$ as a function of the normalized isospin chemical potential
$\mu_I/m_{\pi}$ at $T =$ $\mu_B$ $= 0$.}
\label{s3 1}
\end{figure}
\begin{figure}
\includegraphics[width=3.5in]{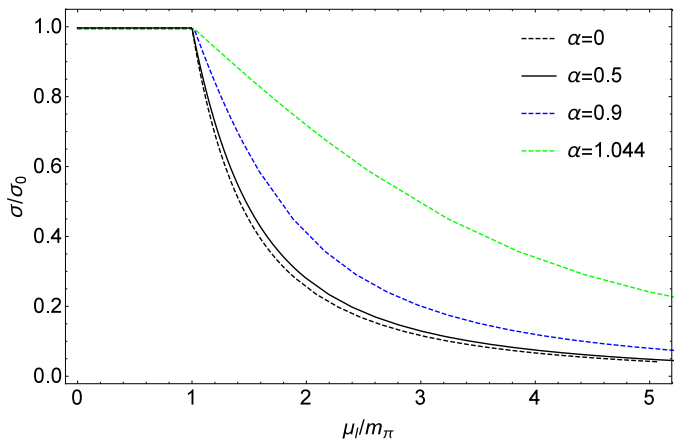}
\caption{The normalized quark condensate $\sigma/\sigma_0$ as a function of the normalized isospin chemical potential
$\mu_I/m_{\pi}$ at $T =$ $\mu_B$ $= 0$.}
\label{s3 2}
\end{figure}

By solving the Eq. (\ref{s2 4}) and Eq. (\ref{s2 5}) simultaneously the condensates $\pi$ and $\sigma$ scaled by the quark condensate
$\sigma_0=-2(250 \text{MeV})^3$ in the vacuum with different $\alpha$'s at $T =$ $\mu_B$ $= 0$ are shown in Fig. \ref{s3 1} and
Fig. \ref{s3 2} respectively. We can see from Fig. \ref{s3 1} that the pion condensates with different $\alpha$'s all keep the vacuum value
(i.e. $\pi = 0$ corresponding to the system in the normal phase) at smaller isospin chemical potential, and then they all go up when
$\mu_I >m_{\pi}$, and the larger the value of $\alpha$, the smaller the pion condensate. The largest difference of pion condensate
with these different $\alpha$'s occurs at $\mu_I \sim 1.5m_{\pi}$. Note that when the critical isospin chemical potential equals to the pion
mass ($\mu_I^c=m_{\pi}$) there is the onset of pion condensation (i.e. $\pi\neq 0$ corresponding to the system in the pion superfluidity phase) and the critical isospin chemical potential $\mu_I^c$ is not changed as $\alpha$ changes. As shown in Fig. \ref{s3 2}, we have similar behavior for the $\sigma$ condensate, the $\sigma$ condensates with different $\alpha$'s all keep the vacuum value $\sigma_0=-2(250 \text{MeV})^3$ at first, and after that they all go down as $\mu_I$ increases.

In the following, we will compare our results for some thermodynamic quantities with the corresponding recent Lattice QCD results \cite{B}.
Due to only the pressure $p$ and the energy density $\epsilon$ relative to the physical vacuum $\Omega_v=\Omega_{(T=\mu_I=\mu_B=0)}$,
which in the mean field approximation we have
\begin{eqnarray}
\label{s3 7}
(\Omega_v)_m &=& \frac{(M_{(T=\mu_I=\mu_B=0)}-m_0)^2}{4G} - 4N_c \nonumber\\
& & \int_{0}^{\Lambda} {d^3 {p}\over (2\pi)^3} \sqrt{|\mathbf{p}|^2 + M^2_{(T=\mu_I=\mu_B=0)}}~,
\end{eqnarray}
can be measured, we introduce the rescaled thermodynamic potential
\begin{equation}
\label{s3 8}
\Omega_r = \Omega - (\Omega_v)_m.
\end{equation}
The energy density $\epsilon$ at zero temperature and zero baryon density and finite isospin density is defined as
\begin{equation}
\label{s3 9}
\epsilon = -p + \mu^{\prime}_I n_I ~~~~(p=-\Omega_r).
\end{equation}

\begin{figure}
\includegraphics[width=3.5in]{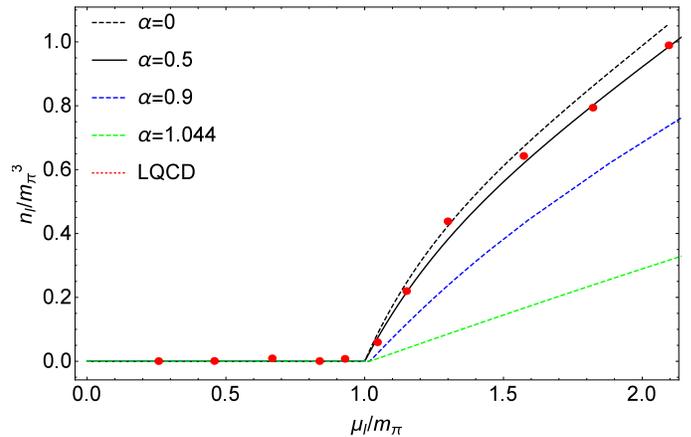}
\caption{The normalized isospin density $n_I/m_{\pi}^3$ as a function of the normalized isospin chemical potential $\mu_I/m_{\pi}$ at $T =$ $\mu_B$ $= 0$.}
\label{s3 3}
\end{figure}
\begin{figure}
\includegraphics[width=3.5in]{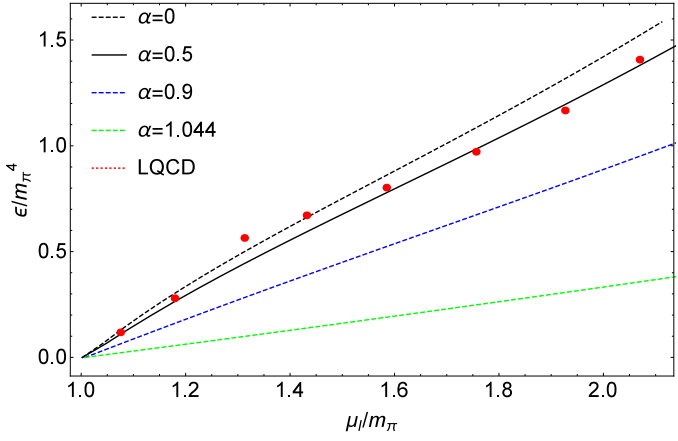}
\caption{The normalized energy density $\epsilon/m_{\pi}^4$ as a function of the normalized isospin chemical potential $\mu_I/m_{\pi}$ at $T =$ $\mu_B$ $= 0$.}
\label{s3 4}
\end{figure}

In Figs. \ref{s3 3} and \ref{s3 4}, the normalized isospin density and energy density with respect to the isospin chemical potential
scaled by $m_{\pi}$ are shown, respectively. These plots have mainly focussed on the region of $\mu_I \lesssim 2m_{\pi}$ throughout which
lattice QCD data are available. Generally, the maximum value of $\mu_I$ within Lattice QCD calculations is constrained by the value of the
lattice spacing. From the figures we can see that the Lattice QCD data can be described well by our calculation with $\alpha=0.5$, although
some data are located on the $\alpha=0$ curve (i.e. the standard NJL model results) around $\mu_I \sim 1.5m_{\pi}$. More lattice data for
isospin density can clarify the situation. These results mean the contributions of the vector channels play an important role in isospin medium.

\begin{figure}
\includegraphics[width=3.5in]{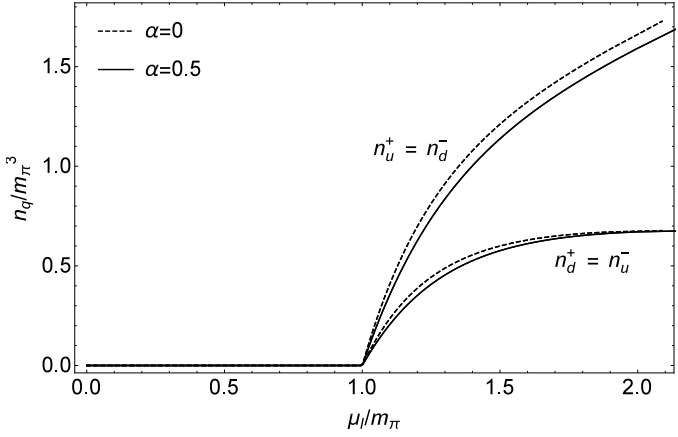}
\caption{The normalized flavor densities $n_q/m_{\pi}^3$ as a function of the normalized
isospin chemical potential $\mu_I/m_{\pi}$ at $T =$ $\mu_B$ $= 0$.}
\label{s3 5}
\end{figure}

The nonzero net isospin density $n_I=(n_u-n_d)/2$ in the superfluidity phase $(\pi\neq 0)$ in Fig. \ref{s3 3} is caused
by the Bose-Einstein condensate of charged pions which leads to different flavor densities. In order to see this clearly,
we plot the flavor densities for each flavor with $\alpha=0$ and $\alpha=0.5$ in Fig. \ref{s3 5}. Each pure flavor number density can be
expressed from the difference between the corresponding quark number density and antiquark number density,
$n_u = n^+_u-n^-_u, n_d = n^+_d-n^-_d$. Making use of the method of the positive and negative energy projectors \cite{He},
$n^{\pm}_{u,d}$ can be separated from $n_{u,d}$,
\begin{eqnarray}
\label{s3 10}
n_u^+ &=& N_c\int_{0}^{\Lambda} {d^3 \vec{p} \over (2\pi)^3} \left[f(E_p^-)+f(-E_p^-) \right.\ \nonumber\\
      &+& \frac{E_p-\mu^{\prime}_I/2}{E_p^-} \left.(f(E_p^-)-f(-E_p^-))\right]\ ,\\
n_u^- &=& -N_c\int_{0}^{\Lambda} {d^3 \vec{p} \over (2\pi)^3} \left[f(E_p^+)+f(-E_p^+) \right.\ \nonumber\\
      &-& \frac{E_p+\mu^{\prime}_I/2}{E_p^+} \left.(f(E_p^+)-f(-E_p^+))-2\right]\ ,\\
n_d^+ &=& N_c\int_{0}^{\Lambda} {d^3 \vec{p} \over (2\pi)^3} \left[f(E_p^+)+f(-E_p^+) \right.\ \nonumber\\
      &+& \frac{E_p+\mu^{\prime}_I/2}{E_p^+} \left.(f(E_p^+)-f(-E_p^+))\right]\ ,\\
n_d^- &=& -N_c\int_{0}^{\Lambda} {d^3 \vec{p} \over (2\pi)^3} \left[f(E_p^-)+f(-E_p^-) \right.\ \nonumber\\
      &-& \frac{E_p-\mu^{\prime}_I/2}{E_p^-} \left.(f(E_p^-)-f(-E_p^-))-2\right]\ .
\end{eqnarray}

From Fig. \ref{s3 5}, one gets the relation $n_u^+ = n_d^- > n_d^+ = n_u^-$ for both $\alpha=0$ and $\alpha=0.5$, which results in
the net isospin density. Therefore the number of $\pi^+$ in the system should be larger than the number of $\pi^-$. Besides, we find that
the flavor number density $n_u^+ (=n_d^-)$ with $\alpha=0.5$ is smaller than that with $\alpha=0$ in the superfluidity phase, and the same
behavior is obtained for $n_u^- (=n_d^+)$, but the difference between $\alpha=0$ and $\alpha=0.5$ of $n_u^- (=n_d^+)$ is smaller than
that of $n_u^+ (=n_d^-)$. From this we can see that the difference between $\alpha=0$ and $\alpha=0.5$ of the nonzero net isospin density
mainly comes from the contribution of $n_u^+ (=n_d^-)$, i.e. the difference in the number of $\pi^+$.

\begin{figure}
\includegraphics[width=3.5in]{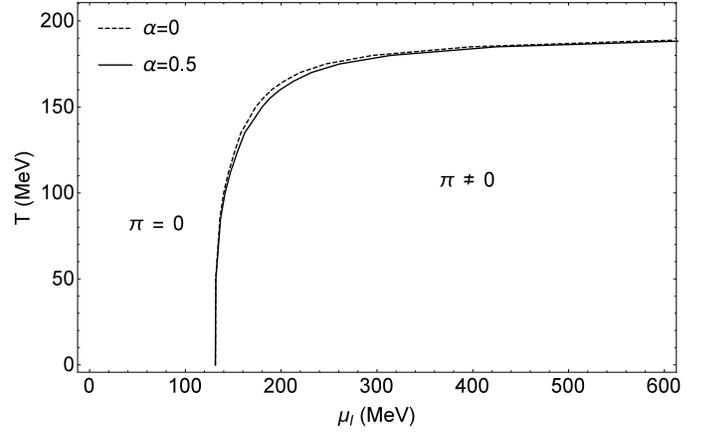}
\caption{The phase diagram in $T-\mu_I$ plane $(\mu_B=0)$.}
\label{s3 6}
\end{figure}

The solution of Eq. (\ref{s2 5}) for $\pi$ separates the region of the pion superfluidity phase ($\pi\ne 0$) from the region of the normal
phase ($\pi=0$). The phase transition lines with $\alpha=0$ and $\alpha=0.5$ delimitating these two regions are given in Fig. \ref{s3 6} in
the $T-\mu_I$ plane for $\mu_B=0$. It can be seen that in the real world with nonzero current quark mass, the system for $\alpha=0$ and
$\alpha=0.5$ are both in the pion superfluidity phase at high enough temperature, regardless of the value of the isospin chemical potential;
But at low temperature, the system are both in the normal phase only at low isospin chemical potential, especially at zero temperature the
critical isospin chemical potential $\mu_I^c$ of the phase transition both have $\mu_I^c=m_{\pi}$.

As depicted in Fig. \ref{s3 6}, the phase transition line with $\alpha=0$ is located above the one with $\alpha=0.5$ in the $T-\mu_I$ plane.
So for the same isospin chemical potential the critical temperature of the phase transition with $\alpha=0.5$ is smaller than that of
$\alpha=0$. In another word, at the fixed isospin chemical potential, the temperature for the occurrence of the phase transition in the
case $\alpha=0.5$ is lower than that in the case $\alpha=0$, and at $\mu_I \sim 1.5m_{\pi}$ the difference between $\alpha=0$ and $\alpha=0.5$
is the largest, which is up to $5\%$.

\section {Summary}
\label{s4}
In this paper, the self-consistent mean field approximation of NJL model is employed to study the contributions of the vector channels
in the finite isospin chemical potential. A free parameter $\alpha$, which reflects the weight of different interaction channels, is
introduced in the model. In our calculation, we consider the contributions of the scalar, vector (not appear in the standard Lagrangian
of NJL model \cite{S.P.Klevansky}) and pseudoscalar-isovector channels with different cases of $\alpha=0$, $\alpha=0.5$, $\alpha=0.9$,
and $\alpha=1.044$, and compare relevant thermodynamic quantities with the recent Lattice QCD data \cite{B}. We find that in superfluid
phase $(\pi\neq 0)$, the lower value of the pion condensate and the higher value of the $\sigma$ condensate appear with the increasing
$\alpha$ compared with that of the $\alpha=0$ case (the standard NJL model), and the largest difference of the pion condensate occurs
at $\mu_I \sim 1.5m_{\pi}$. Our results show that, when $\alpha=0.5$ the isospin density and energy density agree with lattice data well
except around $\mu_I \sim 1.5m_{\pi}$. This indicates the contributions of the vector channels in isospin medium. To see the difference
of the isospin density between $\alpha=0$ and $\alpha=0.5$ more clearly, we also plot the different flavor densities versus isospin chemical
potential and then find that the difference mainly comes from the contribution of the flavor density $n_u^+ (=n_d^-)$ (i.e. the difference
in the number of $\pi^+$ between $\alpha=0$ and $\alpha=0.5$). Finally, we draw the phase diagram in the $T-\mu_I$ plane for $\mu_B=0$ and
show that for the fixed isospin chemical potential with the temperature increasing the occurrence of the phase transition in the case
$\alpha=0.5$ is early than the case $\alpha=0$ and also at $\mu_I \sim 1.5m_{\pi}$ the difference becomes the largest which is up to $5\%$.
In conclusion, through considering the contributions of the vector channels ($\alpha=0.5$) in our study we can get the results (including
the isospin density and energy density) which match lattice QCD data better compared with the standard NJL model ($\alpha=0$) except
around $\mu_I \sim 1.5m_{\pi}$ and this means that the vector channels play an important role in isospin medium.

\acknowledgments
This work are supported by National Natural Science Foundation of China (under Grants No. 11475085, No. 11535005, 11775118, 11690030 and No. 11905104)
and National Major state Basic Research and Development of China (2016YEF0129300).


\begin{thebibliography}{99}
\bibitem{Boyanovsky} {D. Boyanovsky, H. de Vega, and D. Schwarz, Ann. Rev. Nucl. Part. Sci. {\bf 56}, 441 (2006).}
\bibitem{Pizzone} {R. G. Pizzone, R. Spart\'{a}, M. La Cognata, L. Lamia, C. Spitaleri, C. A. Bertulani, A. Tumino,
  Int. J. Mod. Phys. Conf. Ser. {\bf 49}, 1960012 (2019).}
\bibitem{Bielich} {J. Schaffner-Bielich, Nucl. Phys. A{\bf 835}, 279 (2010).}
\bibitem{Weber} {Fridolin Weber, Prog. Part. Nucl. Phys. {\bf 54}, 193 (2005).}
\bibitem{Marty} {R. Marty and J. Aichelin, Phys. Rev. C{\bf 87}, 034912 (2013).}
\bibitem{Xiaofeng} {Xiaofeng Luo, EPJ Web Conf. {\bf 141}, 04001 (2017).}
\bibitem{Tannenbaum} {M.J. Tannenbaum, Conference: C{\bf 16-06-14.2}, 395-414 Proceedings.}
\bibitem{Schmidt} {M. Schmidt, Ukr. J. Phys. {\bf 64}, 640 (2019).}
\bibitem{XiaofengLuo} {Xiaofeng Luo, Nu Xu, Nucl. Sci. Tech. {\bf 28}, 112 (2017).}
\bibitem{Stefano} {Stefano Carignano, Luca Lepori, Andrea Mammarella, Massimo Mannarelli, Giulia Pagliaroli, Eur. Phys. J. A{\bf 53}, 35 (2017).}
\bibitem{D.T.Son} {D.T. Son, Misha A. Stephanov, Phys. Rev. Lett. {\bf 86}, 592-595 (2001).}
\bibitem{Andrea} {Andrea Mammarella, Massimo Mannarelli, Phys. Rev. D{\bf 92}, 085025 (2015).}
\bibitem{Migdal} {A. B. Migdal, E. Saperstein, M. Troitsky, and D. Voskresensky, Phys. Rept. {\bf 192}, 179¨C437 (1990).}
\bibitem{Kogut} {J.B. Kogut, D. Toublan, Phys. Rev. D{\bf 64}, 034007 (2001).}
\bibitem{B} {Bastian B. Brandt, Gergely Endr$\ddot{o}$di, Eduardo S. Fraga, Mauricio Hippert, Jurgen Schaffner-Bielich, Sebastian Schmalzbauer, Phys. Rev. D{\bf 98}, 094510 (2018).}
\bibitem{Viktor} {Viktor Begun, Wojciech Florkowski, Phys. Rev. C{\bf 91}, 054909 (2015).}
\bibitem{Karsch} {F. Karsch, Lect. Notes Phys. {\bf 583}, 209-249 (2002).}
\bibitem{Son} {D.T. Son, Misha A. Stephanov, Phys. Atom. Nucl. {\bf 64}, 834-842 (2001).}
\bibitem{Loewe} {M. Loewe, C. Villavicencio, Phys. Rev. D{\bf 67}, 074034 (2003).}
\bibitem{Klein} {B. Klein, D.Toublan, J. J. M. Verbaarschot, Phys. Rev. D{\bf 68}, 014009 (2003).}
\bibitem{Arai} {R. Arai, N. Yoshinaga, Phys. Rev. D{\bf 78}, 094014 (2008).}
\bibitem{Barducci} {A., R.Casalbuoni, G.Pettini, L.Ravagli, Phys. Rev. D{\bf 69}, 096004 (2003).}
\bibitem{He} {Lianyi He, Pengfei Zhuang,Phys. Lett. B{\bf 615}, 93-101 (2005).}
\bibitem{Tao} {Tao Xia, Lianyi He, Pengfei Zhuang, Phys. Rev. D{\bf 88}, 056013 (2013).}
\bibitem{Lian} {Lian-yi He, Meng Jin, Peng-fei Zhuang, Phys. Rev. D{\bf 71}, 116001 (2005).}
\bibitem{S.P.Klevansky} {S. P. Klevansky, Rev. Mod. Phys. {\bf 64}, 649 (1992).}
\bibitem{Walecka} {J. D. Walecka, Ann. of Phys. {\bf 83}, 491 (1974); B.D. Serot and J.D. Walecka, Adv. Nucl. Phys. {\bf 16}, 1 (1986).}
\bibitem{S} {S.-S. Xu, Z.-F. Cui, B. Wang, Y.-M. Shi, Y.-C. Yang, and H.-S. Zong, Phys. Rev. D {\bf 91}, 056003 (2015).}
\bibitem{BW} {B. Wang, Y.-L. Wang, Z.-F. Cui, and H.-S. Zong, Phys. Rev. D {\bf 91}, 034017 (2015).}
\bibitem{YL} {Y. Lu, Z.-F. Cui, Z. Pan, C.-H. Chang, and H.-S. Zong, Phys. Rev. D {\bf 93}, 074037 (2016).}
\bibitem{ZF} {Z.-F. Cui, I.-C. Clo$\ddot{e}$t, Y. Lu, C. D. Roberts, S. M. Schmidt, S.-S. Xu, and H.-S. Zong, Phys. Rev. D{\bf 94}, 071503 (2016).}
\bibitem{Yang} {Li-Kang Yang, Xiaofeng Luo, Hong-Shi Zong, Phys. Rev. D{\bf 100}, 094012 (2019).}
\bibitem{Kunihiro} {T. Kunihiro and R. Hatsuda, Prog. Theor. Phys. {\bf 74}, 765 (1985).}
\bibitem{FeiWang} {Fei Wang, Yakun Cao, Hongshi Zong, Chin. Phys. C{\bf 43}, 084102 (2019).}
\bibitem{TongZhao} {Tong Zhao, Wei Zheng, Fei Wang, Cheng-Ming Li, Yan Yan, Yong-Feng Huang, Hong-Shi Zong, Phys. Rev. D{\bf 100}, 043018 (2019).}
\bibitem{Q} {Qingwu Wang, Chao Shi, Hong-Shi Zong, Phys. Rev. D{\bf 100}, 123003 (2019).}
\bibitem{J.I.Kapusta} {J. I. Kapusta and C. Gale, Finite-temperature field theory: Priciples and applications (Cambridge University Press, 2006).}
\bibitem{Brandt} {B. B. Brandt and G. Endr\"{o}di, Private communications.}
\end{thebibliography}
\end{document}